\documentclass[reprint,amsmath, twocolumn, amssymb, aps, notitlepage, superscriptaddress, prl]{revtex4-2}

\usepackage{amsmath,amsfonts}    % need for subequations
\usepackage{graphicx}   % need for figures
\usepackage{verbatim}   % useful for program listings
\usepackage{color,xcolor,soul}      % use if color is used in text
\usepackage{subfigure}  % use for side-by-side figures
\usepackage[colorlinks=true, citecolor=blue, urlcolor=blue]{hyperref}
\usepackage{textcomp}
\usepackage{bm}
\usepackage{bbm}
\usepackage{braket}
\usepackage[normalem]{ulem}
\usepackage{natbib}
\sethlcolor{yellow}
\newcommand{\be}{\begin{equation}}
\newcommand{\ee}{\end{equation}}
\newcommand{\ba}{\begin{array}}
\newcommand{\ea}{\end{array}}
\newcommand{\bqa}{\begin{eqnarray}}
\newcommand{\eqa}{\end{eqnarray}}

\graphicspath{{figs/}}

\definecolor{gr}{RGB}{225,225,225}

\newcommand{\hsd}[1]{{\color[rgb]{.9,.1,.1}{#1}}}

\newcommand\hil[1]{\colorbox{yellow}{\textcolor{red}{#1}}}

\renewcommand{\selectlanguage}[1]{}%to avoid language select error for a book entry

\begin{document}

% \title{Optimal Qubit Storage in Inhomogeneous Spin Ensembles in Lossy Cavity}
\title{Optimally Controlled Storage of a Qubit in an Inhomogeneous Spin Ensemble}

\author{Rahul Gupta}
% \email[]{rahul.quantumfield@iitb.ac.in}
\affiliation{Department of Physics, Indian Institute of Technology Bombay, Powai, Mumbai 400076, India}
\affiliation{Physics Department, Blackett Laboratory, Imperial College London, Prince Consort Road, SW7 2AZ, United Kingdom}

\author{Florian Mintert}
\affiliation{Physics Department, Blackett Laboratory, Imperial College London, Prince Consort Road, SW7 2AZ, United Kingdom}

\author{Himadri Shekhar Dhar}
% \email[]{himadri.dhar@iitb.ac.in}
\affiliation{Department of Physics, Indian Institute of Technology Bombay, Powai, Mumbai 400076, India}
\affiliation{Centre of Excellence in Quantum Information, Computation, Science and Technology, Indian Institute of Technology Bombay, Mumbai 400076, India}

\date{\today}

\begin{abstract}

The storage of quantum information in spin-ensembles is limited by practically unavoidable inhomogeneous broadening, and the macroscopic number of spins in such an ensemble makes the design of control solutions 
% \flo{for the increase of coherence time}
% \sout{for}
to increase the coherence time
%to increase the lifetime
a challenging task.
Together with a concurrently developed Krylov theory that allows us to treat the control problem efficiently, we design optimal
% \sout{driving patterns}
{cavity modulation} for such spin ensembles that achieve an {order of magnitude enhancement in qubit lifetime compared to the losses due to {inhomogeneity and cavity decay.}}
% cavity decay and inhomogeneity.}
% \sout{, when both are comparable.}  }

% cavity lifetime, even in the regime where inhomogeneity lifetime is close to cavity lifetime.}

% \flo{Imperial affiliation for Rahul}
% \hsd{Needs to be written.} We present a general scheme for storing a photonic qubit in a macroscopic inhomogeneous spin ensemble in a lossy cavity. Under usual conditions, the quantum information is prone to decoherence because of both the inhomogeneity and cavity loss, however, we present a floquet control protocol that enables the extended storage time, countering both of these effects. Our protocol can be easily applied to superconducting qubits, NMR systems and NV centers.
%%
\iffalse
\hil{We present a theoretical formalism} for storing a photonic qubit in an inhomogeneous spin ensemble. The stored information is prone to significant loss as excitations leak to the larger Hilbert space due to the intrinsic inhomogeneity of the spins and also decay to the cavity vacuum. Using Krylov basis states to effectively represent the exact information dynamics, a periodic cavity modulation scheme is designed that successfully protects the information from loss and enables extended storage time. The protocol can be readily applied to quantum memories based on solid state or atomic ensembles coupled to superconducting or optical resonators.
\fi

\end{abstract}

% \pacs{}% insert suggested PACS numbers in braces on next line

\maketitle 

\noindent\emph{Introduction}.-- 
Over the years, several quantum computing protocols~\cite{Brion2007,Saffman2008,Tordrup2008,Surmacz2008} have been proposed that encode and process qubits in the collective internal states or spatial modes of an ensemble of identical quantum systems. 
For instance, hybrid quantum systems~\cite{Kurizki2015,Clerk2020} have been used to encode information stored in a transmon qubit on to a spin ensemble using a superconducting line resonator~\cite{Imamoglu2009,Wesenberg2009,Kubo2011}. Similar setups have also been used to design spin ensemble based quantum memories for storing qubits~\cite{Wu2010,Tittel2010,Grezes2014}, based on atomic gas~\cite{Hammerer2010,Sangouard2011}, electron spins~\cite{Wu2010}, nitrogen vacancy centers~\cite{Grezes2014, Julsgaard2013a} and nuclear spins~\cite{Witzel2007}. In addition, coherent transfer and storage of information has been enhanced 
% in these systems 
through collective strong coupling~\cite{Kubo2010, Schuster2010} and cavity protection effect~\cite{Putz2014,Putz2017}.

A major limitation in spin ensemble based quantum computing or quantum memory protocol is inhomogeneous broadening~\cite{Gross1982,Julsgaard2013b}. Most spins or emitters in an ensemble are not truly identical, which means that each spin can have slightly different properties such as transition frequencies or variation in coupling with a quantum cavity. The inhomogeneity can give rise to spin dephasing~\cite{Kurucz2011}, leading to loss of any encoded information even in cryogenic setups using high-Q cavity or highly coherent spins~\cite{Putz2014}. Ultimately, these losses need to be mitigated by control techniques such as dynamic decoupling~\cite{deLange2010} and spin refocusing~\cite{Julsgaard2013a}, or by engineering the ensemble using spectral hole-burning~\cite{Putz2017} or as atomic frequency combs~\cite{Riedmatten2007,Afzelius2009, Redchenko2025}. 
However, these protocol assume validity of the semiclassical mean-field approximation,
or they rely on sophisticated and expensive engineering of the spin ensemble.

%{However, these protocols work by either controlling the mean-field, semiclassical dynamics of the ensemble or rely on sophisticated and expensive engineering of the spin ensemble.}
% \flo{What are the limitations of these control techniques? %How do we overcome them?}

% \flo{Designing accurate control protocols requires the ability to simulate the dynamics of the spin ensemble interacting with the cavity field in the deep quantum regime.}
Designing accurate control protocols requires the ability to simulate the dynamics of the spin ensemble interacting with the cavity field in the deep quantum regime.
From a theoretical perspective, 
% studying the dynamics and controlling a qubit encoded in an inhomogeneous spin ensemble 
{this can be challenging for an inhomogeneous ensemble}. Firstly, the inhomogeneity breaks the permutational symmetry of the ensemble, which limits the ability to study the system analytically~\cite{Chase2008}, and secondly, quantum information stored in lower excitation states quickly spreads to the large Hilbert space of the spins, prohibiting accurate numerical solutions (cf.~\cite{Shammah2018}). A quick recourse is to study the system in  the mean-field regime~\cite{Wesenberg2011,Julsgaard2012,Krimer2019}, where spin-echo~\cite{Julsgaard2013a,Julsgaard2013b} and controlled reversible~\cite{Kraus2006} techniques are proposed for quantum memory operations. For small {number of spins} $N$, 
% \flo{$N$ is not defined yet}, 
% beyond mean-field, 
computational studies using cumulant expansion~\cite{Zens2019,Zhang2022} and tensor-network methods~\cite{Dhar2018,Zens2021,Tiwary2025} can be used to simulate the dynamics of information stored in an spin-ensemble. {For large ensembles ({$N \sim 10^3 - 10^{16}$}), the exact dynamics of the system 
% under a typical spin-photon interaction is not 
remains largely intractable beyond mean field.}  

% \flo{we should mention that we consider modulating the cavity frequency.}
% %%
% \flo{In this work, we derive optimized storage protocols for inhomogeneous spin ensembles with storage times that exceed the cavity lifetime \rg{$2/\gamma$} and the \rg{inhomogeneous} broadening time \rg{$\sqrt{2}/\sigma$ by 8 and 15 times respectively. (obtained by fitting fidelity envelope with exponential function for the uniform and optimal case, and with a Gaussian function for the free evolution case)} 
% %\bf XY.
% We overcome the limitations imposed by the macroscopicity of spin ensembles in terms of a Krylov basis, that identifies the collective states into which quantum information leaks.
% While truncation of a Krylov basis is typically a short time approximation, the optimized protocol ensures that the treatment in terms of a truncated Krylov basis remains accurate for the full evolution time.}

In this work, we overcome the limitations imposed by the mere dimension of the Hilbert space in terms of a Krylov representation~\cite{Gupta2026b}. 
% \flo{(we should cite Krylov constrcution)} representation.
This allows us to design optimized protocols to store a qubit in an inhomogeneous spin ensemble, with storage times more than an order of magnitude larger than the lifetime governed by the free ensemble dynamics. While truncation of a Krylov basis is typically valid for short time approximations, the optimized protocol 
with control exerted in terms of a time-dependent modulation of the cavity frequency that does not cause additional excitations,
ensures that the 
dynamics of the system remains accurate for the full evolution time.
%
% \iffalse
%
% {In this work, 
% % we derive 
% an optimized protocol to store a qubit in an inhomogeneous spin ensemble is derived, with storage times more than an order of magnitude larger than the lifetime governed by the free ensemble dynamics.
% This is achieved by coupling the spin ensemble to a cavity and designing an optimal control based on \hsd{time-dependent modulation of} the cavity frequency. Unlike driving based optimal control, 
% % our 
% \hsd{the proposed} protocol does not add any excitations in the Hilbert space -- the information is constrained in the encoded subspace.
% %%
% However, to overcome the limitations imposed by the large, macroscopic number of spins, 
% % we use 
% an effective representation in the Krylov basis \hsd{is used} that
% identifies the collective states into which quantum information leaks. While truncation of a Krylov basis is typically valid for short time approximations, the optimized protocol ensures that the 
% % treatment in terms of a truncated Krylov basis 
% dynamics of the system remains accurate for the full evolution time.
% %%
% \fi
%
In the
%smaller, effective Hilbert 
{truncated Krylov} space, 
% we use Floquet theory~\cite{Shirley1965}, 
% we derive a time-independent Hamiltonian, which allows us 
% to find 
{the optimal period for the cavity frequency modulation is obtained using Floquet theory~\cite{Shirley1965}
% , which 
to achieve} maximum fidelity for a given broadening of the spin frequency distribution and collective spin-cavity coupling. 
%%%
This allows us to 
% optimize the controlled modulation to obtain a pulsed cavity field, which 
extend the storage time of the encoded information for periods much 
% \flo{longer}\sout{larger} 
longer than the losses due to inhomogeneity, as well as the lifetime of the coupled cavity.
% Our
{The} protocol is not specific to any specific spin ensemble and 
can be applied to a variety of hybrid quantum platforms~\cite{Kurizki2015,Clerk2020}.%}

\noindent\emph{Model and setup}.--
The dynamics of an inhomogeneous ensemble of spins or two-level emitters interacting with a single-mode cavity (shown in Fig.~\ref{fig:fid_setup}) is captured by the Tavis-Cummings (TC) model~\cite{Tavis1968}
%,
%where
{Hamiltonian}
\begin{eqnarray}
\mathcal{H}=\omega_c a^\dagger a + \sum_{j=1}^{N} \omega_j \sigma_j^+\sigma_j^- 
% \textrm{and}~\mathcal{H}_I = 
+\sum_{j=1}^{N} g_j (a^\dagger \sigma_j^- + a \sigma_j^+)\ , \label{Ham}
\end{eqnarray}
where $a$ and $a^\dagger$ are the annihilation and creation operator of the cavity with frequency $\omega_c$.
The transition frequency and the spin-cavity coupling of the $j^{th}$ spin in the ensemble 
is $\omega_j$ and $g_j$, respectively. The spin operators
$\sigma^\pm_j = \frac{1}{2}(\sigma^x_j\pm i\sigma^y_j)$, 
% \flo{index $j$}, 
where $\sigma^x_j$ and 
% $\sigma^+_j$ and 
$\sigma^z_j$ are the typical Pauli operators. 
%%
% \sout{For the TC model, the total number of excitations is conserved, i.e. $N_{\rm ex}=\langle a^\dag a\rangle + \langle\sum_{j} \sigma_j^+\sigma_j^-\rangle$ is fixed at all times.
%%
% For a single qubit encoded in the ensemble, }
{The total number of excitations $N_{\rm ex}= a^\dag a + \sum_{j} \sigma_j^+\sigma_j^-$ is conserved,
{and the most relevant dynamics for purposes of information storage takes place in the single-excitation subspace.}
%and the work focuses on states with at most one excitation.}

% We want to work in the low excitation regime, particularly when 
% \sout{there is only a single excitation present in the system, $N_{\rm ex}=1$. }
% for the $j^{th}$ spin in the ensemble.
% a single spin labeled $j$ of the spin ensemble.
% The non-interacting part contains an energy-offset chosen for convenience, such that the eigen-energies of each spin are given by $0$ and $\omega_j$ as opposed to the common convention of $-\omega_j/2$ and $\omega_j/2$.

% While the cavity resonance frequency $\omega_c$ can typically be determined   high accuracy, the mere number of spins makes determination of the spin resonance frequencies $\omega_j$ and the spin-cavity coupling constants $g_j$ practically impossible.
\begin{figure}[t]
    \includegraphics[width=\columnwidth]{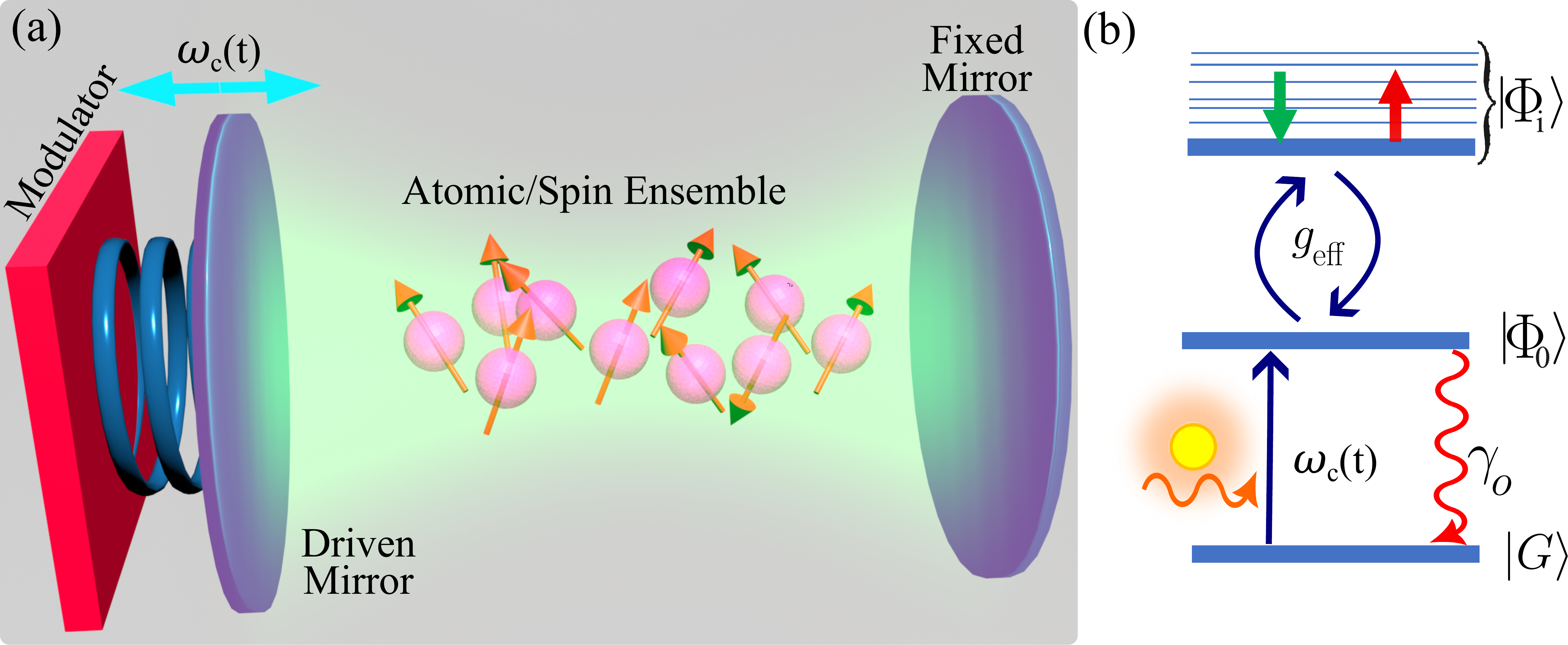}
    \caption{(a) A spin ensemble is kept inside a cavity whose frequency is controlled via a modulator. (b) An effective energy level diagram in a
    joint basis of spins and cavity spanned by $\{\ket{G},\ket{\Phi_i}\}$,
    % Krylov basis , 
    where $\ket{G} = \ket{0}_c \otimes \ket{0}$ is the ground or zero-excitation state and $g_{\rm eff}$ is the spin-cavity coupling.}
    \label{fig:fid_setup}
\end{figure}

For encoding a photonic qubit~\cite{Kok2007} $|\psi\rangle = \alpha|0\rangle_c+\beta|1\rangle_c$ in the spin ensemble, where $|0\rangle_c$ and $|1\rangle_c$ are the vacuum and single-photon Fock states,
the desired spin states are the collective ground state $|0\rangle$ and the single-excitation bright state $|1\rangle$, given by  
\begin{equation}
   |0\rangle = |g\rangle^{\otimes N};~ 
   % |1\rangle = \sum_j g_j P_j (|e\rangle \otimes |g\rangle^{\otimes N-1}),
   % |1\rangle = \sum_{j=1}^N g_j (|g\rangle^{\otimes j-1}\otimes|e\rangle \otimes |g\rangle^{\otimes N-j}),
   |1\rangle = {\sum_{j=1}^N \frac{g_j}{g_{\rm{eff}}} \sigma_j^+|g\rangle^{\otimes N},}
\end{equation}
% \flo{do we really want to define an unnormalized state?
%We define $g_{eff}$ anyway.}
where $g_{\rm{eff}}=\sqrt{\sum_j g_j^2}$ is the effective ensemble coupling to a single photon, $|g\rangle$ and $|e\rangle$ are the ground and excited states of a single spin. 
%%
% \flo{
Given the spatial dependence of the cavity field, there is typically a non-trivial distribution of coupling constants, {\it i.e.} $g_j\neq g_k$ for $j\neq k$, so that the bright state does not necessarily coincide with 
% the 
the symmetric Dicke state~\cite{Dicke1954}. 
This type of disorder is not a concern, because an excitation stored in the bright state can always be converted into an excitation in the cavity.
% }
As such, for a homogeneous ensemble in spin frequencies, $\omega_j = \omega_s$ 
{(for all $j$)}, any information {can be coherently transferred back and fourth between the two states $|0\rangle_c \otimes |1\rangle$ and $|1\rangle_c \otimes |0\rangle\}$ that span the bright part of the single-excitation subspace.}
%stored in the photonic qubit 
%is coherently transferred to the collective spin qubit, \hil{with the single-excitation space spanned by $\{|0\rangle_c \otimes |1\rangle, |1\rangle_c \otimes |0\rangle\}$.}
Inhomogeneity in the resonance frequencies $\omega_j$, on the other hand, 
% \rg{removes the degeneracy and excitations can 
leads to leakage of excitations from the bright state {into the dark part of the single-excitation subspace.}
%to the larger $N-1$ dimensional ``dark'' single-excitation subspace.
% }}

%For a homogeneous ensemble, $\omega_j = \omega_s$ and $g_j = g$, $\forall~j$, the bright state is the symmetric Dicke state~\cite{Dicke1954} \flo{why do we distinguish between the bright state and the Dicke state?}, and any information stored in the photonic qubit is coherently transferred to the collective spin qubit, with the degenerate single-excitation space spanned by $\{|0\rangle_c \otimes |1\rangle, |1\rangle_c \otimes |0\rangle\}$. This is no longer the case in the presence of inhomogeneity. For different $g_j$, the bright state is not symmetric and different $\omega_j$ removes the degeneracy and excitations can leak from the bright state to the larger $N$ dimensional ``dark''` single-excitation subspace.

% \noindent\emph{Optimally controlled storage.}--
{\noindent\emph{Optimal storage.}--
For a qubit encoded in the collective states of the spin ensemble, there are two dominant processes that are detrimental to the stored information. 
% affect the dynamics and subsequently the ability to control and retrieve the information. 
%%
First, while the spin ensemble is strongly coupled to the cavity,
% $g_{\rm{eff}}\gg\sigma$, 
with effective coupling $g_{\rm{eff}}$, the information is coherently stored in a superposition of the cavity and the bright mode of the ensemble. As the qubit is partly stored in the cavity, it is subject to decay that deteriorates the information.
On the other hand, in the limit of large detuning or the dispersive regime, where $\Delta = |\omega_c - \bar{\omega}|\gg g_{\rm{eff}}$ {(with the weighted average spin frequency $\bar{\omega} = \sum g_j^2\omega_j/g_{\rm eff}^2$)}, the cavity and the ensemble are effectively decoupled. 
%Here,  $\bar{\omega} = \sum g_j^2\omega_j/g_{\rm eff}^2$ is the weighted average spin frequency.
In this regime, the information is far from cavity losses, but  
leaks to the single-excitation dark subspace due to inhomogeneity. 
% \rg{We plan to overcome these limitations by enabling time dependent control over the cavity frequency $\omega_c$.}
%%%
% \flo{There is a big jump in the argument ... in particular, we have not said that we consider time-dependent cavity frequency}
% \hsd{I think it is not necessary at this point -- as one could have done this using other mechanisms and also the first part just needs strong coupling.}  
%This leak can be protected by
% \rg{The leakage to dark subspace}

% \flo{
{The leakage into the dark subspace can be suppressed with a phase modulation of the bright state that results in destructive interference in the dark subspace.}
% \sout{The leakage to the dark subspace 
% can be protected by introducing 
% {is {suppressed} if there is} a phase shift of $\pi$ between the initial and the time-evolved state, which results in destructive interference.}
% between the bright and dark modes.
%\flo{not sure if modes can interfere ... can we and after 'interference'?}
A rapid phase 
% \flo
{modulation could} 
% \sout{shift can} 
be achieved through Rabi oscillations when the spin ensemble is resonantly coupled to the cavity ($\Delta = 0$). However, this comes at the cost of information loss due to cavity decay. As such, an ideal protocol would involve {time dependent control over the cavity frequency $\omega_c$, with
periodic modulation 
% of cavity frequency $\omega_c$, 
to alternate between the dispersive ($\Delta(t) \gg 0$) and resonant ($\Delta(t) = 0$) regimes.} A critical advantage of such a protocol over conventional control based on external driving is that no additional excitations are introduced, and the system always remains in the single-excitation or qubit subspace.} 

%{Let us consider
% \flo{
Such a protocol can be realized with} a periodic, piecewise constant
% \sout{, time-dependent }
detuning 
% \flo{is it really piecewise constant?}\rg{-yes it is indeed piecewise constant taking fixed values of either 0 or $\Delta$, now updated clearly in the following eqn.}
$\Delta(t) = \omega_c(t) - \bar{\omega}$, such that
\begin{equation}
\Delta(t)=\begin{cases}
\Delta &\textrm{for}~t+nT\in[0,t_0/2)\cup[t_0/2 +t_\pi,T)\\
0 &\textrm{for}~t+nT\in[t_0/2,t_0/2+t_\pi),
\end{cases}
\label{eq:protocol}
\end{equation}
with the period $T=t_0+t_\pi$ consisting of off-resonant ($\Delta\gg g_{\rm{eff}}$) duration $t_0$ and resonant $\pi$ pulse duration $t_\pi=\pi/g_{\rm{eff}}$.
The ratio between resonant and dispersive dynamics is %determined
{parametrized} by $t_0$, which then governs the optimal storage and retrieval of information.
% \begin{eqnarray}
% \Delta(t) = 0\ &\mbox{ for }&\ t+nT\in[t_i,t_i+t_\pi]~\textrm{and} \nonumber \\
% \Delta(t) \gg g_{\rm{eff}}\ &\mbox{ for }&\
% t+nT\in[t_i+t_\pi,t_i+T],
% \label{eq:protocol}
% \end{eqnarray}
% with the period $T$ and an initial time $t_i$. 
% Experimental advances have enabled different 
Such detunings can be experimentally achieved by cavity frequency modulation that can be applied to a variety of spin ensembles~\cite{Wilson2011,Pasi2013,Xu2020}.

\begin{figure*}[htb]
    \includegraphics[width=2\columnwidth]{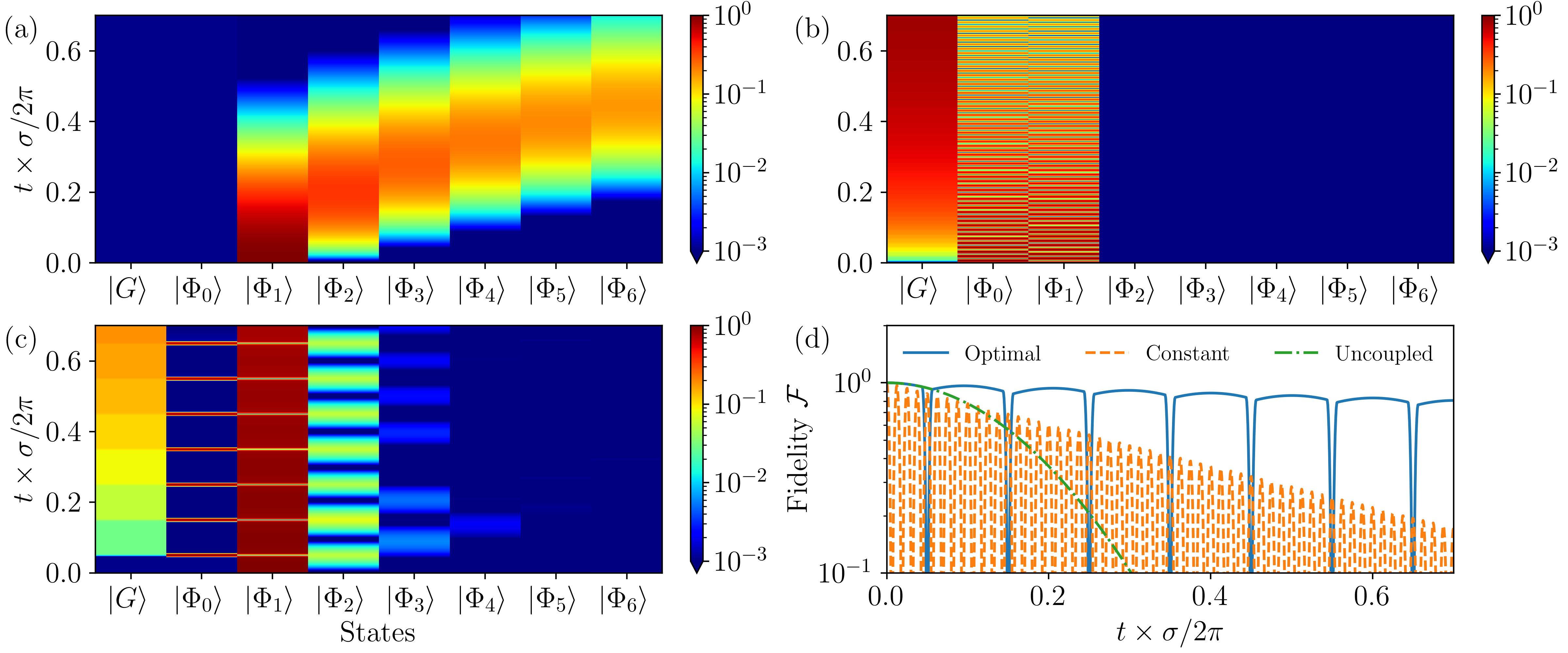}
    \caption{Loss and retrieval of information in a Gaussian spin ensemble.
    The plots (a)-(c) shows the fidelity of the time-evolved state with the states $\vert i \rangle=\{\vert G\rangle, \vert \Phi_0\rangle,\vert \Phi_1\rangle,...\}$, where $\vert G\rangle=\vert0\rangle_c\otimes\vert g\rangle^{\otimes N}$ and $\{\vert\Phi_j\rangle\}~\forall j\in]\{0,M-1\}$ are the Krylov states.  %\flo{(the identification $\ket{i+1}$ with $\ket{\phi_i}$ is a bit confusing)}
    The plots represent fidelity (a) in the absence of a coupled cavity, (b) for  an unmodulated cavity, and (c) optimally modulated cavity with {period
    $T= 0.1\times2\pi/\sigma$.}
    % $t_0\simeq 0.1~2\pi/\sigma$. 
    % with time in periods of $\sigma$ for spin excitation in (a) absence of cavity coupling, (b) constant cavity coupling, and (c) optimally controlled cavity coupling. 
    Plot (d) shows the fidelity of the bright state $\vert \Phi_1\rangle$ for the above 3 cases. 
    The effective coupling strength used here is $g_{\rm eff}=50~\sigma$ with cavity dissipation rate $\gamma=\sigma$ and spin ensemble width $\sigma/2\pi$. Higher fidelity, close to unity, is achieved for either $\gamma\rightarrow 0$ or $g_{\rm{eff}} \gg \sigma,\gamma$. All axes are dimensionless.}
    % \comm{@Rahul: Can we add a plot in End matter comparing $\mathcal{F}$ with $t_0$ and $\sigma/g_{\rm eff}$.}
    % \flo{We get an improvement compared to 'uncoupled' around $t\simeq 0.1\sigma/2\pi$, but then the fidelity if reduced to about $0.9$. Is there a limiting case (fast, strong driving) in which we maintain perfect fidelity?
    % Can we plot fidelity at the point in time in which fidelity is maximal as function of $\sigma$ and $\gamma$, or as function of $t_0$ and $\Delta$?
    % In Fig. 1c one can not see wether occupations leak beyond the state $\ket{\phi_2}$; can we also plot on logscale (maybe for appendix) }\rg{-limiting cases where perfect Fidelity is retrieved are when $g_{\rm{eff}}\gg\sigma,\gamma_0$ (even for a strong $\gamma_o$, small $\sigma$ enhances storage greatly). Unit Fidelity is also retrieved as $\gamma_o\rightarrow0$, for an indefinitely long time.}}
    \label{fig:fid_comp}
\end{figure*}

% \noindent\emph{Optimal information dynamics.}--
\noindent\emph{Floquet based control.}-- 
% \flo{
Given the macroscopic number of spins in a spin ensemble and the correspondingly macroscopic number of resonance frequencies, it is essential that a control protocol does not rely on detailed values of the individual resonance frequencies $\omega_j$.
{The framework of a Krylov basis allows us to express the system Hamiltonian only in terms of statistical moments of the distribution of resonance frequencies.
The concurrently developed framework with a detailed derivation can be found in Ref.~\cite{Gupta2026b}.}
%A framework that expresses the system Hamiltonian only in terms of statistical moments of the distribution of resonance frequencies can be found in Ref.~\cite{Gupta2026b}.
%the {\bf what the right expression}.

{The Krylov basis includes the two states
$\ket{\Phi_0}=|1\rangle_c \otimes |0\rangle$ and $\ket{\Phi_1}=|0\rangle_c \otimes |1\rangle$ as well as states $\ket{\Phi_j}$ with $j>1$ resultant from orthogonalization of the states
$\mathcal{H}\ket{\Phi_{j-1}}$.}
\iffalse
\hsd{Consider a suitably defined Krylov basis $\{\ket{\Phi_p}\}$, where $\ket{\Phi_0}=|1\rangle_c \otimes |0\rangle$ and $\ket{\Phi_1}=|0\rangle_c \otimes |1\rangle$ are the 
% =\{|1\rangle_c \otimes |0\rangle,|0\rangle_c \otimes |1\rangle\}$
single photon and bright state in the joint basis.}
% },
%{\hil{To implement} the optimal protocol, the dynamics of information encoded in the spin ensemble \hil{is studied} using an effective, orthonormal Krylov basis $\{\ket{\Phi_p}\}$~\cite{Gupta2026b}, which is numerically more tractable than the large $N$ dimensional single-excitation subspace.
%
\fi
%{Now,
{For a Gaussian distribution of the spin frequencies $\omega_j$ and standard deviation $\sigma=\langle(\omega-\bar{\omega})^2\rangle/g^2_{\rm{eff}}$, 
% the central moments read
% \begin{equation}
% C_p=\begin{cases}
%     \sigma^p(p-1)!!~~\mbox{for even values of $p$, and}\\
%     0~~~~~~~~~~~~~~\mbox{for odd values of $p$, }
% \end{cases}
% \end{equation}
% with the $(p-1)!!=\prod_{k=1}^{p/2}(2{k}-1)$, where $p$ is even.
% % \flo{even/odd??}\rg{-p is even here, formulae updated for clarity}. 
% From Ref.~\cite{Gupta2026b}, 
the elements of the Hamiltonian $\mathcal{H}$ in the Krylov basis are
% in Eq.~\eqref{Ham} are 
% In this case, the only finite matrix elements read
% $\bra{\Phi_p}{\cal H}\ket{\Phi_p} = \omega_c$ for $p=0$ and $\bar{\omega}$ for $p\neq 0$, while }
% \hsd{@Rahul: What is $\sigma$ here?}\rg{-defined now}
\begin{eqnarray}
\bra{\Phi_p}{\cal H}\ket{\Phi_p}&=&\begin{cases}
\omega_c &\textrm{for}~~p=0\\
\bar{\omega} &\textrm{for}~~p\neq 0
\end{cases},~\textrm{and}\nonumber\\
% \end{eqnarray}
% \begin{eqnarray}
\bra{\Phi_p}{\cal H}\ket{\Phi_{p+1}} &=& \bra{\Phi_{p+1}}{\cal H}\ket{\Phi_p} =\sqrt{p}~\sigma ~~\textrm{for}~p\neq0,
% \bra{\Phi_p}{\cal H}\ket{\Phi_{p+1}} &=& \bra{\Phi_{p+1}}{\cal H}\ket{\Phi_p} =\begin{cases}
%     g_{\rm{eff}} &\textrm{for}~~p=0\\
%     \sqrt{p}~\sigma &\textrm{for}~~p\neq0
% \end{cases}
% ,~~\textrm{where}
% \nonumber\\
% &\sigma^2=C_2 = \sum_jg_j^2(\omega_j-\bar\omega)^2{/g^2_{\rm{eff}}}. &%= (\langle\omega^2\rangle - \langle\omega^2\rangle)/g^2_{\rm{eff}}.&
\label{eq:sqrt_origin}
\end{eqnarray}
with $\bra{\Phi_{0}}{\cal H}\ket{\Phi_1}=\bra{\Phi_{1}}{\cal H}\ket{\Phi_0}=g_{\rm{eff}}$.
To describe the controlled dynamics under cavity modulation, the time-dependent $\mathcal{H}$, with  $\omega_c(t) = \bar{\omega} + \Delta(t)$, can be transformed under the unitary $U(t)$, such that 
$\mathcal{H}_m =U^{\dag}\mathcal{H}U-iU^{\dag}{\partial U}/{\partial t}$. The explicit form of $U(t)$ and the subsequent transformation to find the modulated Hamiltonian is shown in Sec.~I of the End Matter.
In the Krylov basis, the modulated Hamiltonian is given by $\mathcal{H}_m(t) = \mathcal{H}_m^s + \mathcal{H}_{m}^{\rm int}(t)$:
\begin{align}
    \mathcal{H}_m^s&=\sum_{j=1}^{M-1}\sqrt{j}\sigma\left(\vert\Phi_j\rangle\langle\Phi_{j+1}\vert + \vert\Phi_{j+1}\rangle\langle\Phi_j\vert\right)~\textrm{and}\label{Ham:mod_spin}\\
    \mathcal{H}_{m}^{\rm int}(t) &= g_{\rm{eff}}\left(e^{i\varphi(t)}\vert\Phi_0\rangle\langle\Phi_1\vert + e^{-i\varphi(t)}\vert\Phi_1\rangle\langle\Phi_0\vert\right),
    \label{Ham:mod_int}
\end{align}
where $\varphi(t) =\int_{0}^t dt'\Delta(t')$.
% \flo{Do we really need this? Interaction picture is fairly standard, and we have lots of technical apsects to discuss.}\rg{-Here we by this representation, we wanted to show two things, firstly, the representation of Full Hamiltonian in Krylov basis, which isn't explicitly done till this point, secondly we wanted to emphasize that any effect of time-dependent detuning around cavity frequency comes as a time-dependent phase in the interaction term only, so basically one can tweak $g_{\rm{eff}}(t)$ by modulating cavity frequency, for example, in Eq.~\eqref{eq:large_detuning}, we see that a large detuning can effectively switch off the interaction. Also, later on, the piecewise constant nature of detuning helped us to write the Floquet Hamiltonian.}
%%
% \noindent\emph{Floquet dynamics.}-- 
%%%

{The optimal cavity modulation can be found with Floquet theory~\cite{Shirley1965}.}
\iffalse
{To find the optimal cavity modulation, 
% \hil{we study} 
the 
% The stroboscopic 
time-evolution of the hybrid system under the periodic, time-dependent Hamiltonian \hsd{is studied}
% can be studied 
using Floquet theory~\cite{Shirley1965}. 
\fi
%Specifically, 
Since the Hamiltonian $\mathcal{H}_m(t)$ in Eqs.~\eqref{Ham:mod_spin} and \eqref{Ham:mod_int} is periodically time-dependent, the stroboscopic propagator can be expressed as 
\begin{align}
    e^{-iH_F T}:=\mathcal{T}\left[\exp\left(\int_{0}^{T}\mathcal{H}_m(t) dt\right)\right]
\end{align}
in terms of a time-independent Hamiltonian $H_F$. {This effective Hamiltonian} can be expressed using Floquet-Magnus expansion~\cite{Magnus1954,Kuwahara2016} as $H_{F}=\sum_{k=0}^{\infty}T^k H^{(k)}_F$, 
% \hsd{(perhaps we should cite Magnus' original paper here)}
where the zeroth and first order terms are given by
\begin{align}
H^{(0)}_F&=\mathcal{H}^s_m + \frac{g_{\rm{eff}}}{T}\left(\vert\phi_0\rangle\langle\phi_1\vert\int_0^T e^{i\varphi(t)}dt + \textrm{h.c.} \right),\nonumber\\
&=\mathcal{H}^s_m + \frac{g_{\rm{eff}}t_{\pi}}{T}\hat{\sigma}^x_{01},~\textrm{and}\label{first-order}
\end{align}
\begin{align}
H^{(1)}_F
% &=\frac{\sigma}{2iT^2}\int_0^T dt_1\int_0^{t_1}dt_2\left(-ig_{\rm{eff}}(t_2)\hat{\sigma}^y_{02}+ig_{\rm{eff}}(t_1)\hat{\sigma}^y_{02}\right),\nonumber\\
&=\frac{i\sigma\hat{\sigma}^y_{02}}{2iT^2}\int_0^T dt_1\left(g_{\rm{eff}}(t_1)t_1 -\int_0^{t_1}g_{\rm{eff}}(t_2)dt_2\right)dt_1,\nonumber\\
    % H^{(1)}_F 
&=\frac{g_{\rm eff}\sigma\hat{\sigma}^y_{02}t^2_\pi}{8T^2},
% \right.\nonumber\\
% \nonumber\\
% &\left.+ \vert\phi_1\rangle\langle\phi_0\vert\int_0^T e^{-i\varphi(t)}dt\right).
\label{second-order}
\end{align}
where $\hat{\sigma}^y_{ki}=-i(\vert \Phi_k\rangle\langle \Phi_i \vert - \vert \Phi_i\rangle\langle \Phi_k \vert)$. Therefore, the effective, time-independent Floquet Hamiltonian up to first-order is given by 
\begin{equation}
H_F = \mathcal{H}^{s}_m + \frac{g_{\rm{eff}}t_{\pi}}{T}\hat{\sigma}^x_{01} + \frac{g_{\rm{eff}}\sigma t^2_{\pi}}{8T}\hat{\sigma}^y_{02}.
\label{floq-eff}
\end{equation}
See Sec.~II of End Matter for derivation of Eqs.~\eqref{first-order}-\eqref{floq-eff}.

The optimal controlled dynamics of the system can now be examined in terms of the periodic evolution under $H_F$. If the system is initially in the bright state {$\ket{\Phi_1}$}, the stroboscopic fidelity at every period $T$ is given by $\mathcal{F}_{n}=\vert\langle\Phi_1\vert e^{-inH_F T}\vert\Phi_1\rangle\vert^2$. 
{Numerically optimizing this over a single period $T$
% , for a Gaussian spin ensemble with standard deviation $\sigma$, 
gives optimal fidelity $\mathcal{F}_{n=1}$ at 
$T\simeq 0.1~T_\sigma + t_\pi$, where $T_\sigma=2\pi/\sigma$ 
and $\sigma$ is the standard deviation 
% \flo
{of the distribution of resonance frequencies defined above Eq.~\eqref{eq:sqrt_origin}}.
This gives the optimal value of $t_0$ in Eq.~\eqref{eq:protocol} as $t_0 = 0.1~T_\sigma$.}

\noindent\emph{Numerical study of controlled dynamics.}--
Figure~\ref{fig:fid_comp} shows the dynamics of excitation in the {bright state $\vert\Phi_1\rangle$}. 
% \flo{we also use bright state and we use two different symbols $\ket{1}$ and $\ket{\Phi_1}$)}  of the spin ensemble.} 
In the absence of a cavity, the information quickly disperses to the higher Krylov states $\{\ket{\Phi_i}\}$, as shown in Fig.~\ref{fig:fid_comp}(a). This highlights the loss of information due to the inhomogeneity in the spin ensemble. 
When coupled to a resonant cavity, with no modulations ($\Delta(t) = 0$) and an effective coupling strength $g_{\rm{eff}}$, the system undergoes Rabi oscillations, with the excitation coherently transferred to the cavity state. In the presence of cavity losses, the system rapidly decays to the ground state $|0\rangle$ or the no-excitation subspace as the cavity decays to the vacuum with a loss rate $\gamma_0$.
This is shown in Fig.~\ref{fig:fid_comp}(b). 
In the presence of a numerically optimized cavity modulation, as described in Eq.~\eqref{eq:protocol}, 
% \sout{note that} 
the spread of the information to the higher Krylov states as well to the ground state is well contained, and the excitation remains in the bright mode over several cycles, as shown in Fig.~\ref{fig:fid_comp}(c). 
The strength of the controlled dynamics is captured by the loss of fidelity between the time-evolved state and the initial excited state, for all the three cases, {as shown in Fig.~\ref{fig:fid_comp}(d).
%%%
% Figure~\ref{fig:fid_comp}(d) shows that the time-evolved state have \hil{significantly high} fidelity with the initial state under the controlled modulation. 
Using the modulation period $T$ 
% = 0.1 \times  2\pi/\sigma$ 
as the unit of time, it is observed that fidelity $\mathcal{F}$ for both 
% without or 
uncoupled and
unmodulated cavity drops below $1/2$ by $t=2T$. However, for modulated cavity, the fidelity $\mathcal{F} = 0.8$ at $t=7T$. Using a numerical fit, the fidelity lifetime for the encoded information is calculated to be around $34T$, which is 
% \sout{more than} \rg
{around} 20 times larger than the lifetime due to inhomogeneity ($1/\sigma$) and cavity decay ($1/\gamma$).}
% Figure~\ref{fig:fid_comp}(d) shows that the time-evolved state have \hil{significantly high} fidelity with the initial state under the controlled modulation. 
The simulation of the dynamics of the system under cavity losses is governed by the master equation defined in Sec.~III of End Matter. 

%Figure~\ref{fig:fid_theta} shows the fidelity of different encoded initial states.
% \rg{The change of slope of Fidelity evolution from negative to positive value can be seen with each resonant $\pi$ pulse extending the survival of bright mode.
%The quantum Zeno \cite{Itano1990} type of effect can be seen with each resonant $\pi$ pulse extending the survival of bright mode.

% Our next goal here is to store a photonic qubit state $\vert\psi_0(\theta,\phi)\rangle=\sin(\theta/2) \vert 0\rangle_c + \cos(\theta/2) e^{i\phi}\vert 1\rangle_c$ with arbitrary superposition amplitudes and relative phases $(\theta,\phi)$ between vacuum state $\vert 0\rangle_c$ and single phton fock state $\vert 1 \rangle_c$, into the inhomogeneous spin ensemble for a prolonged time.

\begin{figure}[t]
    \includegraphics[width=\columnwidth]{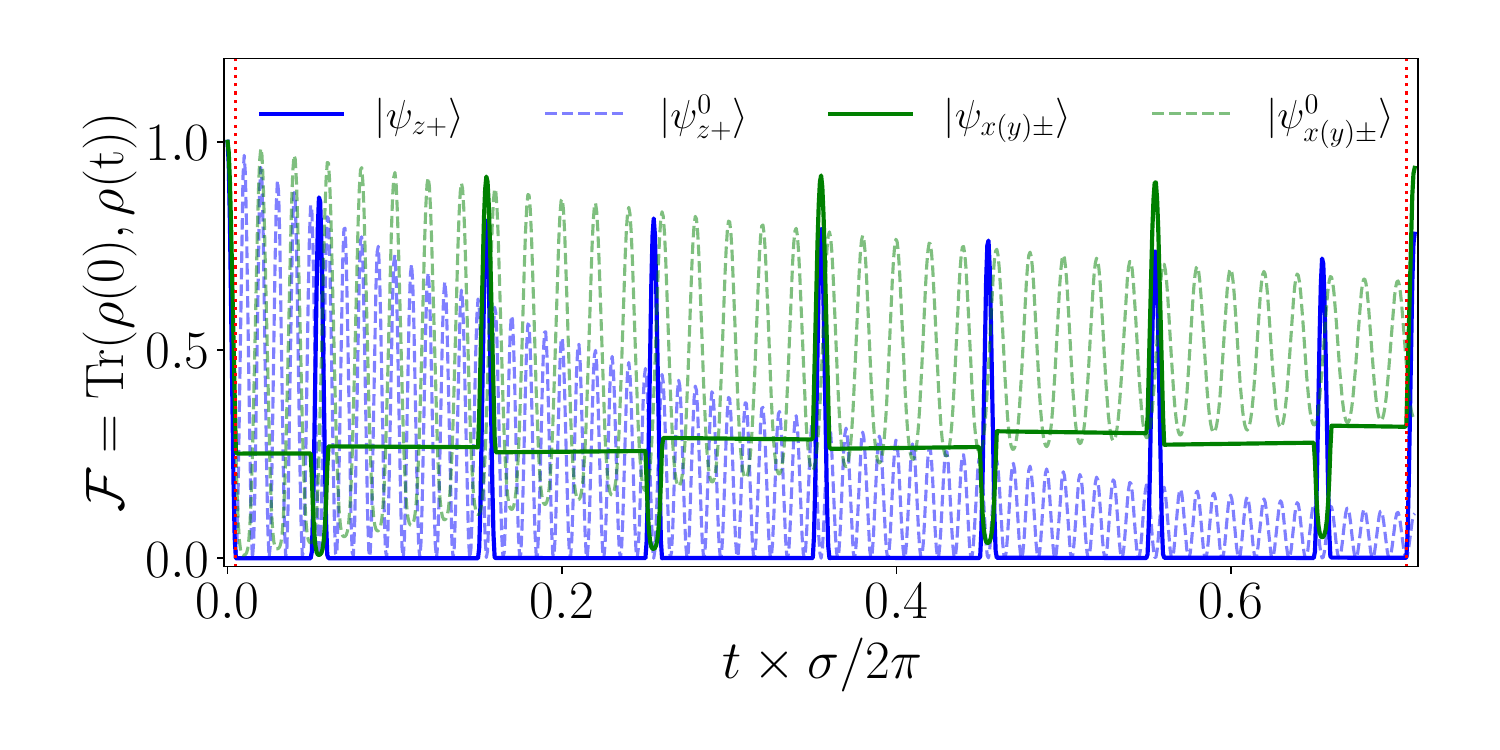}
    \caption{Time evolution of photonic qubits, initialized as eigenstates $\vert\psi_{j\pm}\rangle$ of Pauli matrices $\sigma_j,~\forall j\in\{x,y,z\}$ in the Fock state basis $\{\vert 0 \rangle_c, \vert 1 \rangle_c\}$. The solid lines show the temporal change of fidelity of states $\vert\psi_{j\pm}\rangle$, under cavity modulation, while the dashed lines show the fidelity for states $\vert\psi_{j\pm}^0\rangle$, with no modulation. Both axes are dimensionless.}
    % Here, $\vert\psi_{x(y)\pm}\rangle$ are the eigenstates of $\sigma_{x(y)}$ and have identical evolution while $\vert\psi_{z+}\rangle$ is the eigenstate of $\sigma_{z}$ follows different evolution. Fidelity is independent of the relative phase between the states and only depend on superposition amplitudes.}
    \label{fig:fid_theta}
\end{figure}

{To highlight the efficacy with which a qubit can be encoded in the spin ensemble, the input photonic qubit can be initialized as one of the eigenstates of the Pauli operators $\sigma_j$ for $j\in \{x,y,z\}$, given by $|\psi_{j\pm}\rangle$. {Here, $|\psi_{z\pm}\rangle$ correspond to the $|1\rangle_c$ and $|0\rangle_c$ states.} 
%%%
Figure~\ref{fig:fid_theta} shows the fidelity of the photonic state as the hybrid system evolves, where the unmodulated case is shown by the states $|\psi_{j\pm}^0\rangle$.
The fidelity of all superposition states are {higher than that of {$\ket{\psi_{z+}}$}}, which maps to the bright state $\ket{\Phi_1}$. Thus, optimization of the modulation for the bright mode ensures that all other qubit states are optimally encoded in the ensemble and are stroboscopically retrievable. Note that eigenstates of $\sigma_{x(y)}$ have identical fidelity during the evolution. As such, the control protocol is independent of the relative phase and only depends on coefficients of the qubit.}

\noindent\emph{Discussions.}-- 
% In this work, we 
% \sout{The work proposes an elegant optimal control 
% design an optimal control 
% protocol based on modulation of the cavity frequency that allows 
% us to encode and store 
% encoding and storage of a qubit in an inhomogeneous spin ensemble.} 
The 
% \flo
{present} 
protocol has two key advantages over known methods -- first, by modulating the cavity frequency any additional excitations are avoided in the hybrid setup, unlike control protocols based on external drives, and secondly, representing the system in the Krylov basis allows 
% \flo
{for}
% us to exactly 
{the exact study of the information} dynamics in a much smaller Hilbert space. 
The optimally controlled storage is applicable to quantum information and computing protocols in a host of 
% several 
platforms including Rydberg atoms, solid-state spins and superconducting qubits, with potential for use in quantum state preparation and in designing memory and sensing applications. An extension of the protocol is to consider controlled dynamics of information in higher excitation space that would allow for multiple qubits or qudit encoding and storage, which would allow for greater versatility in its operation.

% \comm{Need to write.}
% \hsd{We show that a single-rail photonic qubit in a cavity, encoded in an inhomogeneous spin ensemble, can be controlled and protected using a physically implementable cavity modulation based on periodic changes in cavity detuning. To find the optimal period we make use of the representation of the hybrid system in an effective Krylov basis, which simplifies the study of the dynamics in terms of the statistical properties  of the ensemble. This allows us to use Floquet theory and optimize the fidelity of the encoded information over several cycles. The formalism can be applied to a wide variety of spin ensemble based quantum computing and quantum memory platforms.}      

\begin{acknowledgments}
% \flo{visiting position}
R.G. acknowledges funding from CSIR-HRDG, India in the form of Senior Research Fellowship and thanks Imperial College London for support during a visit through the Global Development Hub Fellows Fund.
H.S.D. acknowledges financial support from SERB-DST, India via a
Core Research Grant (No: CRG/2021/008918) and the Industrial Research \& Consultancy Centre, IIT Bombay via grant RD/0521-IRCCSH0-001 (No: 2021289).
\end{acknowledgments}

\bibliography{Bibliography_file}

\clearpage

\appendix
\section*{End Matter}

% \setcounter{equation}{0}
% \noindent \textit{Appendix A: 
% \noindent \textit{Appendix B: 

\subsection{I. Hamiltonian under cavity modulation\label{Em_1}}

To find the modulated Hamiltonian in the Krylov basis, we start with the time-dependent Hamiltonian $\mathcal{H}(t)$, i.e. the Hamiltonian in Eq.~\eqref{Ham} but with a time-dependent cavity frequency $\omega_c(t) = \bar{\omega} + \Delta(t)$, given by
\begin{eqnarray}
\mathcal{H}(t)=\omega_c(t) a^\dagger a + \sum_{j=1}^{N} \omega_j \sigma_j^+\sigma_j^- 
% \textrm{and}~\mathcal{H}_I = 
+\sum_{j=1}^{N} g_j (a^\dagger \sigma_j^- + a \sigma_j^+). \nonumber\\
\end{eqnarray}
Now, to describe the dynamics under cavity modulation, $\mathcal{H}$ can be transformed under the unitary $U(t)$, where
$\mathcal{H}_m =U^{\dag}\mathcal{H}U-iU^{\dag}{\partial U}/{\partial t}$, and
% . The explicit form of $U(t)$ and the subsequent transformation to find the modulated Hamiltonian is shown in Sec.~I of the End Matter.
% To describe this, we write down the time-dependent Hamiltonian by plugging $\omega_c(t)$ into $\mathcal{H}$ and perform a unitary transformation
\begin{align}
    U(t)&=\exp{\left(-i(\bar{\omega}_c a^\dag a + \sum_{j=1}^N \bar{\omega} \sigma_j^+\sigma_j^-)t - i\varphi(t) a^{\dag}a\right)},\nonumber\\
    &\textrm{with}~\varphi(t)=\int_{0}^t dt'\Delta(t').
    \label{app:ut}
\end{align}
For resonant case ${\omega}_c=\bar{\omega}$, the transformation gives the {modulated Hamiltonian
\begin{align}
% \mathcal{H}_m&=U^{\dag}\mathcal{H}U-iU^{\dag}\frac{\partial U}{\partial t},\\
\mathcal{H}_m&=\sum_{j=1}^N\Delta_j\sigma^+_j\sigma^-_j + g_{\rm{eff}}\left(e^{i\varphi(t)}a^\dag\Sigma^- + e^{-i\varphi(t)}a\Sigma^+\right),\label{eq:H_m_coll}
\end{align}
where, $\Sigma^{\pm}=\sum_j g_j\sigma^{\pm}_j/g_{\rm{eff}}$. Note that the excitation 
number $N_{\rm{ex}}$ is conserved under this transformation.
% This modulation conserves the excitation number $N_{\rm{ex}}$. 
In the single-excitation regime, the modulated Hamiltonian $\mathcal{H}_m$ in Eq.~\eqref{eq:H_m_coll} can be represented in the Krylov basis, 
% thus simplifying it to a time-dependent Krylov space Hamiltonian
%, in the single-excitation regime, we can express interaction terms with collective raisig and lowering operators $\Sigma^{\pm}=\sum_j g_j\sigma^{\pm}_j/g_{\rm{eff}}$ which simplifies the modulated Hamiltonian as
%\begin{align}
%H_m&=\sum_{j=1}^N\frac{\Delta_j}{2}\sigma^z_j + g_{\rm{eff}}\left(e^{i\varphi(t)}a^\dag\Sigma^- + e^{-i\varphi(t)}a\Sigma^+\right)
%\end{align}
\begin{align}
    \mathcal{H}_{m}&=\mathcal{H}_m^s + \mathcal{H}_{\rm{int}}\label{eq:H_m},~\textrm{where}\\
    \mathcal{H}_m^s&=\sum_{j=1}^{M-1}\sqrt{j}\sigma\left(\vert\Phi_j\rangle\langle\Phi_{j+1}\vert + \vert\Phi_{j+1}\rangle\langle\Phi_j\vert\right),~\textrm{and}\\
\mathcal{H}_{\rm{int}}&=g_{\rm{eff}}\left(e^{i\varphi(t)}\vert\Phi_0\rangle\langle\Phi_1\vert + e^{-i\varphi(t)}\vert\Phi_1\rangle\langle\Phi_0\vert\right),
\end{align}
where we write $a^\dag \Sigma^-=\ket{\Phi_0}\langle\Phi_1\vert$. Here, we have used the relations in Eq.~\eqref{eq:sqrt_origin} to obtain the elements in the Krylov basis. Note that the dynamics so far is not dependent on the optimal control or the specific cavity modulation.

% where we have used Eq.~\eqref{eq:sqrt_origin} to express the bare spin Hamiltonian $\mathcal{H}_m^s$ in Krylov basis.
% %\flo{
% The derivation up to this point is independent of the specific driving pattern.

% matrices of size $N_c\times N_c$ and $M\times M$ respectively, where $N_c, M$ are the dimensions of truncated cavity fock space and the Krylov space for ensemble respectively. $\rho$ is the joint density matrix of size $N_c M\times N_c M$. We take $N_c=2,M=128$ ensuring no finite size effect induced revivals, and simulated the dynamics using Qutip python library \cite{lambert2026}.

\subsection{II. Floquet dynamics\label{EM_2}}

In this section, we look at the derivation of the time-independent, Floquet Hamiltonian, up to the first order. As earlier, we express the modulated Hamiltonian Eq.~\eqref{eq:H_m} by performing the Floquet-Magnus (FM) expansion~\cite{Shirley1965,Magnus1954,Kuwahara2016} such that
\begin{align}
    e^{-iH_F T}:=\mathcal{T}\left[\exp\left(\int_{0}^{T}\mathcal{H}_m(t) dt\right)\right],
\end{align}
where $H_{F}=\sum_{k=0}^{\infty}T^k H^{(k)}_F$. The zeroth and first order Hamiltonians are given by
\begin{align}
    H^{(0)}_F&=\frac{1}{T}\int_0^Tdt \mathcal{H}_m(t),~\textrm{and}\\
    H^{(1)}_F&=\frac{1}{2iT^2}\int_0^Tdt_1\int_0^{t_1} dt_2\left[\mathcal{H}_m(t_1),\mathcal{H}_m(t_2)\right].
    % H^{(2)}_F&=\frac{-1}{6T^3}\int_0^Tdt_1\int_0^{t_1} dt_2\int_0^{t_2} dt_3\left(\left[H_m(t_1)\left[H_m(t_2),H_m(t_3)\right]\right]+\left[H_m(t_3)\left[H_m(t_2),H_m(t_1)\right]\right]\right)
\end{align}

\begin{figure}[t]
    \centering
    \includegraphics[width=\columnwidth]{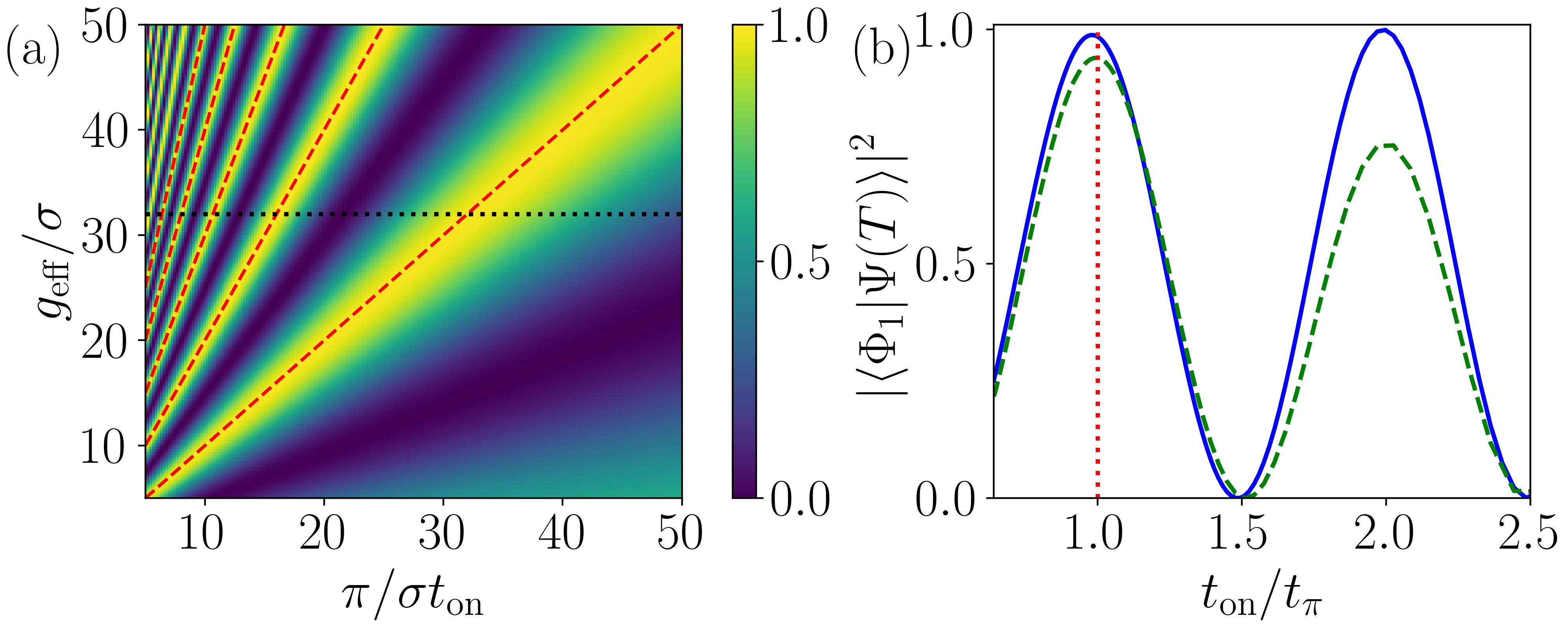}
    \caption{(a) Optimal period $t_{\rm on}$ when cavity is resonant and coupling is switched on: (a) variation of fidelity with $g_{\rm{eff}}$ and $t_{\rm{on}}$ 
    % variation for 
    over a {single period of} evolution under $H_F$. The fidelity is maximum at $t_{\rm{on}}= m \pi/g_{\rm eff}=m t_\pi$ (red dashed lines), where $m$ is an integer. 
    % is maximum under evolution
    % shows optimal $t_{\rm{on}}=mt_\pi=m\pi/g_{\rm{eff}}$, where $m$ \rg{is positive integer}. 
    (b) Dynamics of the system under the master equation (green dashed curve) shows that 
    % beyond $m=1$, 
    fidelity starts decreasing compared to Floquet dynamics (blue solid curve) for $t_{\rm{on}} > t_\pi$. Here, $g_{\rm{eff}}=32\sigma$. All axes are dimensionless.}
    % , thus we fix $t_{\rm{on}}=t_\pi,m=1$.}
    \label{fig:ton_opt}
\end{figure}
The zeroth-order Floquet Hamiltonian is simply obtained by averaging over the Hamiltonian $\mathcal{H}_m(t)$ over a single time period $T$. Here, a single period $T$ consists of a period $t_{\rm{on}}$, where the cavity is resonant ($\Delta(t)=0$) and coupling is switched on, and for time $t_0$ where the detuning is finite ($\Delta(t)=\Delta$) and coupling is effectively switched off 
% detuning for time $t_0$, 
i.e., $T=t_{\rm{on}}+t_0$. 
% For this, 
As such, the phase $\varphi(t)$ in Eq.~\eqref{app:ut} is given by
% \rg{$T=t_{\rm{on}}+t_0$, consisting of resonant detuning for time $t_{\rm{on}}$ and off-resonant detuning for time $t_0$}. For this, the phase $\varphi(t)$ is
\begin{align}
    \varphi(t)&=\begin{cases}
    \Delta t~~&0<t<t_0/2\\
    \Delta t_0/2~~&t_0/2<t<T - t_0/2\\
    \Delta \left(t-t_{\rm{on}}\right)~~&T - t_0/2<t<T
    \end{cases},
\end{align}
which then leads to 
\begin{align}
H^{(0)}_F=\mathcal{H}_m^s &+ \frac{g_{\rm{eff}}}{T}\left(\vert\Phi_0\rangle\langle\Phi_1\vert\int_0^T e^{i\varphi(t)}dt + \textrm{h.c.} \right).
% \right.\nonumber\\
% \nonumber\\
% &\left.+ \vert\phi_1\rangle\langle\phi_0\vert\int_0^T e^{-i\varphi(t)}dt\right).
\end{align}
Now, we evaluate integrals as:
\begin{align}
    \int_0^T e^{\pm i\varphi(t)}dt&=\left(\frac{2}{\Delta}\sin\left(\frac{\Delta t_0}{2}\right) + t_{\rm{on}}\right)e^{\pm i\Delta t_0/2}.
    % \int_0^T e^{-i\varphi(t)}dt
    % &=\left(\frac{2}{\Delta}\sin\left(\frac{\Delta t_0}{2}\right) + t_{\pi}\right)e^{-i\Delta t_0/2}
\label{eq:large_detuning}
\end{align}
Since $\Delta\gg2/t_{\rm{on}},2/t_0$, we can neglect the sinusoidal term in the integral and assume that the phase factor is set to 1. 
% Effectively, the coupling $g_{\rm{eff}}\approx 0$, or is switched off, for the period $t_0$. 
This simplifies the Hamiltonian
% so the duration $t_0$ acts as a switch-off time where coupling $g_{\rm{eff}}\rightarrow0$, which simplifies the result
\begin{align}
    H^{(0)}_F&=\mathcal{H}_m^s + \frac{g_{\rm{eff}}t_{\rm{on}}}{T}\left(\vert\Phi_0\rangle\langle\Phi_1\vert + \vert\Phi_1\rangle\langle\Phi_0\vert\right)\nonumber\\
    &=\mathcal{H}_m^s + \frac{g_{\rm{eff}}t_{\rm{on}}}{T}\hat{\sigma}^x_{01}.
\end{align}

\begin{figure}[t]
    \centering
    \includegraphics[width=\columnwidth]{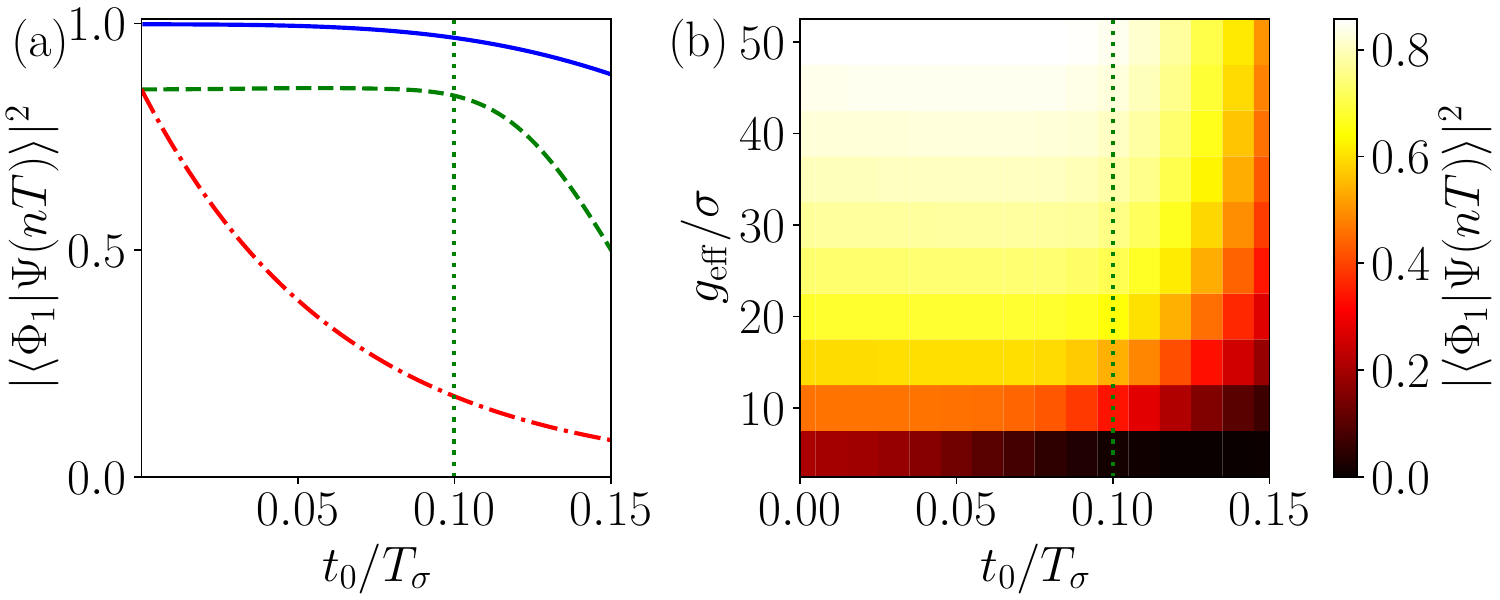}
    \caption{Optimal period $t_0$ when coupling is switched off: (a) fidelity variation during evolution over a single period under $H_F$ (solid blue) and over five periods ($n=5$) under the master equation (ME) (green dashed). The fidelity drops sharply for $t_0 > 0.1~T_\sigma$. The uncoupled free dynamics is shown by red dash-dot line. (b) Fidelity change during dynamics under ME for different $g_{\rm{eff}}$ and $t_0$. The optimal value $t_0 = 0.1T_\sigma$ is shown by a green-dotted line. All axes are dimensionless.}
    % for $g_{\rm{eff}}\gg\sigma$.}
    \label{fig:t0_opt}
\end{figure}

We further verified numerically that under large detuning, this frequency modulation is identical to an effective spin-cavity coupling modulation, where $g_{\rm{eff}}(t)=g_{\rm{eff}}$ during the $\pi$ pulse and $g_{\rm{eff}}(t)=0$ for rest of the period $T$. 
% $g_{\rm{eff}}(t)=0$ for $t_0$ time in a full period $T$ and $g_{\rm{eff}}(t)=g_{\rm{eff}}$ only during the $\pi$ pulse time $t_{\pi}$.
%%
Using the commutation relation,
\begin{align}
 \left[\hat{\sigma}^x_{ij},\hat{\sigma}^x_{jk}\right]&= -i\hat{\sigma}^y_{ki},~ \textrm{where}\nonumber\\
 \hat{\sigma}^y_{ki}&=-i\vert \Phi_k\rangle\langle \Phi_i \vert + i\vert \Phi_i\rangle\langle \Phi_k \vert,
\end{align}
% $\left[\hat{\sigma}^x_{ij},\hat{\sigma}^x_{jk}\right]=-i\hat{\sigma}^y_{ki}$, where  $\hat{\sigma}^y_{ki}=-i\vert \Phi_k\rangle\langle \Phi_i \vert + i\vert \Phi_i\rangle\langle \Phi_k \vert$. 
% With this simplification, we obtain the first-order Floquet Hamiltonian by noting the important commutation relations here as $\left[\hat{\sigma}^x_{ij},\hat{\sigma}^x_{jk}\right]=-i\hat{\sigma}^y_{ki}$, where  $\hat{\sigma}^y_{ki}=-i\vert k\rangle\langle i \vert + i\vert i\rangle\langle k \vert$.
% , thus $\left[\hat{\sigma}^x_{12},\hat{\sigma}^x_{01}\right]=-i\hat{\sigma}^y_{02}$. 
% With this, 
% we obtain 
the first-order Floquet Hamiltonian is
\begin{align}
    H^{(1)}_F&=\frac{\sigma}{2iT^2}\int_0^T dt_1\int_0^{t_1}dt_2\left(-ig_{\rm{eff}}(t_2)\hat{\sigma}^y_{02}+ig_{\rm{eff}}(t_1)\hat{\sigma}^y_{02}\right),\nonumber\\
    &=\frac{i\sigma\hat{\sigma}^y_{02}}{2iT^2}\int_0^T dt_1\left(g_{\rm{eff}}(t_1)t_1 -\int_0^{t_1}g_{\rm{eff}}(t_2)dt_2\right)dt_1,\nonumber\\
    % H^{(1)}_F 
    &=\frac{g_{\rm eff}\sigma\hat{\sigma}^y_{02}t^2_{\rm{on}}}{8T^2}.
    % \implies H_{F}\simeq H^{(0)}_F + TH^{(1)}_F ,\nonumber\\
    % &=H^{\rm{eff}}_s + \frac{g_{\rm{eff}}t_{\pi}}{T}\hat{\sigma}^x_{01} + \frac{g_{\rm{eff}}\sigma t^2_{\pi}}{8T}\hat{\sigma}^y_{02}\label{eq:HF1_derived}
\end{align}
Therefore, up to the first order the effective Floquet Hamiltonian is given by
\begin{align}
H^{\rm eff}_F = H^{(0)}_F + TH^{(1)}_F = \mathcal{H}_m^s + \frac{g_{\rm{eff}}t_{\pi}}{T}\hat{\sigma}^x_{01} + \frac{g_{\rm{eff}}\sigma t^2_{\rm{on}}}{8T}\hat{\sigma}^y_{02}.
\label{eq:HF1_derived}
\end{align}

Thus, once we initialize to the bright mode $\vert\Phi_1\rangle$, the fidelity of getting revival after a single period $T$ of evolution under $H^{\rm eff}_F$ is obtained by numerically optimizing for $t_{\rm{on}}$ and $t_0$ values, for any fixed $g_{\rm{eff}}$. 
% under the time-independent Floquet Hamiltonian numerically. 
{The optimal values for $t_{\rm{on}}$ and $t_0$ are shown in Figs.~\ref{fig:ton_opt} and \ref{fig:t0_opt}, respectively. Fig.~\ref{fig:ton_opt}(a) shows that fidelity is maximum under evolution with $H_F$ for time period $T$ at $t_{\rm{on}}=m t_\pi$, where $m$ is an integer. 
However, the dynamics under the full master equation in Fig.~\ref{fig:ton_opt}(b),  where cavity losses are included, shows that the optimal value of $t_{\rm{on}}$ is for $m=1$ or  $t_{\rm{on}}=t_\pi$. 
% is the only optimal solution. 
Similarly, the optimal time $t_0$, where the coupling is effectively switched off, 
% optimal switch off time by fixing 
for $t_{\rm{on}}=t_\pi$ 
% and varying $t_0$ 
is shown in Fig.~\ref{fig:t0_opt}. Therefore, the optimal values using Floquet theory are $t_{\rm{on}}=t_\pi$ and $t_0\simeq 0.1 T_\sigma$.} \\

\subsection{III. Master equation\label{EM_3}}
To account for losses due to the decay of photons in the cavity, we consider the Lindblad master equation~\cite{Breuer2007}, where the spins are defined in the Krylov basis and the cavity in the Fock basis. The Hamiltonian in the tensor-product space $\mathcal{L}_c \otimes \mathcal{L}_s$ is given by 
% $\gamma_o$, we write a master equation that takes spin Hamiltonian in Krylov form $H_s^{\rm{eff}}$ and use truncated Fock space for cavity photons, as the excitation number is low, assuming negligible thermal photon excitations.
% \hsd{@Rahul: Please make notations consistent}\rg{-done.}
\begin{align}
    \mathcal{H}_m&=I_c\otimes \mathcal{H}_m^s + g_{\rm{eff}}\left(e^{i\varphi(t)}a^{\dag}\otimes \vert\Phi_0\rangle\langle\Phi_1\vert \right.\nonumber\\
    &\left.+e^{-i\varphi(t)}a\otimes \vert\Phi_1\rangle\langle\Phi_0\vert\right),
\end{align}
where $I_{c}$ is the identity operator in the Hilbert space of the cavity $\mathcal{L}_c$. 
Now, the Lindblad master equation can be written as
\begin{align}
    \dot{\rho}&=-i\left[H_{\rm{eff}},\rho\right] + \frac{1}{2}\left(2K_a\rho K^\dag_a - \rho K^\dag_a K_a - K^\dag_a K_a\rho\right),
\end{align}
where the operator $K_a = \sqrt{\gamma}~a\otimes I_s$ denotes cavity decay, with rate $\gamma$. Again, $I_{s}$ is the identity operator in the Hilbert space of the spin ensemble $\mathcal{L}_c$. 
% where $\gamma_0$ is the cavity dissipation rate, $I_c$ and $I_s$ are identity operators. 
For numerical simulations, the dimension of the Krylov basis is taken as $M=128$, with the truncated dimension of the Fock basis as $N_c=2$, which is sufficient for the single-excitation dynamics, in the absence of any incoherent sources or thermal photons. Numerical simulations were partly performed using Qutip Python library \cite{lambert2026}.

\end{document}